\documentclass[5p,twocolumn,number,sort&compress]{elsarticle}
\usepackage[english]{babel}
\usepackage{graphicx}
\usepackage{amsmath,amssymb,amsfonts,bm}
\usepackage{epsfig}
\usepackage{dsfont}
\usepackage{url}
\usepackage{booktabs, makecell}
\usepackage{here}
\usepackage{gensymb}
\usepackage{xcolor}
\usepackage{hyperref}

\renewcommand{\P}{\mathbb{P}} 
\newcommand{\N}{\mathcal{N}} 
\def\R{{\mathds{R}}} 
\newcommand{\GEV}{\text{GEV}} 
\newcommand{\GPD}{\text{GPD}} 

\newcommand{\Wdirm}{W_\text{dir}} 
\newcommand{\Cdirm}{C_\text{dir}} 
\newcommand{\Wdir}{$\Wdirm$} 
\newcommand{\Cdir}{$\Cdirm$} 

\journal{Ocean Engineering}

\begin{document}

\begin{frontmatter}

\title{3-D environmental extreme value models for the tension in a mooring line of a semi-submersible}
  \author[ifr]{N. Raillard\corref{cor1}}
	\author[ifr]{M. Prevosto}
	\author[act]{H. Pineau}
	
	\cortext[cor1]{Corresponding author}
	\address[ifr]{Marine Structures Laboratory, IFREMER, 29280 Plouzan\'e, France}
	\address[act]{Actimar, 36 Quai de la Douane, 29200 Brest}

\begin{abstract}
Design optimization is crucial as offshore structures are exposed to deeper and harsher marine conditions. The structure behaviour is dependent on several joint environmental parameters (wind, wave, currents, \textit{etc.}). Environmental contours are useful representations to provide multivariate design conditions. However, these contours may lead to different design points depending on the method used to compute them, and thus may be misleading to structural engineer.

In this work, we propose to use a response meta-model for the inter-comparison of some state-of-the-art methods available for modelling multivariate extremes, in order to provide a straightforward methodology, focusing on the derivation of three-dimensional contours. The considered case study focuses on the tension in a mooring line of a semi-submersible platform. In a first step, the key met-ocean parameters and the associated load model of the tension in the mooring line are set-up. Several multivariate extreme analysis methods are then applied to derive the environmental contours. {These methods are chosen in order to cover all the possible dependence cases, from extremal dependence to extremal independence.} Conditional Extreme and several extreme value dependence function models are investigated. The physical-space Huseby contouring method is used to derive environmental surface. A comparison with the extreme load extrapolated from the meta-model is provided to assess the performance of each method.
\end{abstract}


\end{frontmatter}

\section{Introduction}

The emergence of floating wind turbines requires an updated design methodology for assessing the environmental conditions. These devices are exposed to the joint loads of wind, wave and currents. { Not} only their intensities but their direction and also the wave frequency influence the design of floating structures. The calculation of the met-ocean extreme values should take into account the dependency between atmospheric and oceanic processes. By studying examples of existing offshore structure in harsh {environments}, our final goal is to refine extreme multivariate analysis in order to reduce the costs and increase the reliability of the floating structures. 

The MulanR (Multivariate Analysis Methodology in function of the Response) project { (see \cite{Pineau2018} for more details)} was launched to get more insight into the state-of-the-art methods available for modelling multivariate extremes, in the aim of better identifying which combinations of { environmental variables} raises extreme responses. 

Many approaches exist to estimate the extreme environment and the related structure response design point. A procedure, called \emph{response based} is based on a numerical model of a structure and may be difficult to apply to complex structures whom behaviour relies complex interactions with the sea-state. The alternative \emph{response independent} approach is based on constructing
the centennial sea-state for the parameters, which is the hyper-surface of equal probability of being exceeded on average once every one hundred years. Then the maximum of the response of the structure is computed on this hyper-surface by the use of a response meta-model. See e.g. \cite{Huseby2015} for a definition of such curves and \cite{Ewans2014} for more precise thoughts about \emph{response based} and \emph{response independent} methods.

The main objective of this study is to inter-compare, by the mean of a response meta-model, the techniques to derive 3-D environmental contours. 

Since most of structure numerical models computation are time consuming, alternative meta-models should be simple and fast-running formula linking the met-ocean variables to the response. The met-ocean parameters to be considered in offshore design include wind speed ($W_s$), current speed ($C_s$), significant wave height ($H_s$) and peak period ($T_p$) of one or more wave systems, as well as the directions of all these components (see Table~\ref{tab_param} for a description of the environmental variables considered in this study).

The building of an environmental joint contour consists in two steps: (i) the statistical dependence modelling and (ii) the contouring method. 

Among the dependence modelling, a perfect dependence hypothesis is compared to several investigated conditional models: the Gaussian Copula (a.k.a. Nataf method), families of extreme value dependence functions (copula), including logistic model (Gumbel copula) and the Conditional Extreme model of Heffernan and Tawn \cite{Heffernan2004}.

Then, we applied the Huseby \cite{Huseby2015} contouring methods which calculates directly the contour in the physical space from Monte Carlo simulations. The work of Huseby has been here extended to 3-D, but extension to higher dimensions is straightforward.

The maximum tension in a mooring line of a semi-submersible in North Sea is considered as an application case for the comparisons of the joint extreme  approaches.{ The introduction of statistical models of structures provides a novel approach to compare extreme values of responses, which are not easily accessible. Related work on the construction of contours taking into account the directionality in the 2D case can be found in \cite{Pineau2018}, along with comparison if the Huseby method to the more classical I-FORM (\cite{Winterstein1993}).}

The first section of this article presents the building of the tension meta-model. Then, the met-ocean components of the extreme environment at site are described. A methodology part reviews the extreme value modelling techniques - both univariate and multivariate - considered in this study and the contouring procedure. The obtained numerical results are presented and inter-compared and some key findings are highlighted.

\begin{table}[ht]
	\centering
		\begin{tabular}{lcl} \toprule
			Name & Symbol & Unit \\ \midrule
			Significant wave height & $H_s$ & meter \\
			Peak wave period & $T_p$ & second \\
			Peak wave direction & $D_p$ & degree \\ 
			Mean wave direction & $D_m$ & degree \\ 
			Wind speed & $W_s$& meter per second \\
			Wind direction & \Wdir & degree \\
			Current speed & $C_s$ & meter per second \\
			Current direction & \Cdir & degree \\ 			\bottomrule
		\end{tabular}
	\caption{Parameters describing the sea-state, symbol used and corresponding units.}
	\label{tab_param}
\end{table}

\section{Meta-model for the mooring line tension}
\label{MetaModel}

\begin{figure}[ht]
	\centering
		\includegraphics[width=\columnwidth]{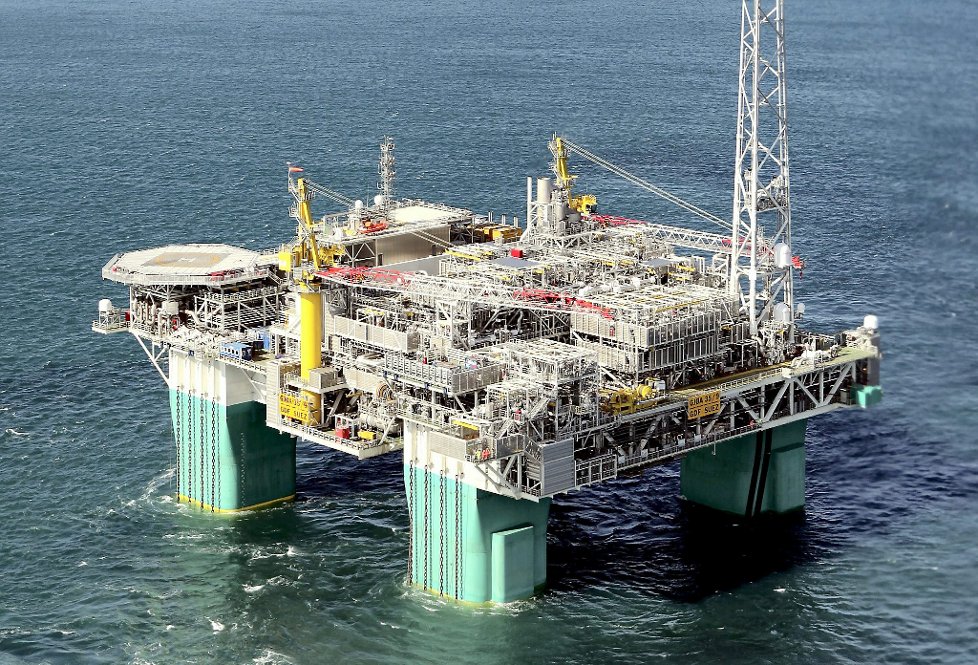}
	\caption{Gj\o a semi-submersible platform (Photo: \O yvind Nesv\r{a}g, source: \url{www.energy-pedia.com})}
	\label{fig:GjoaSemi}
\end{figure}

The Gj\o a semi-submersible is operating on Gj\o a field, an oil and gas field, located 40km West offshore Norway, 100km North of Bergen. The field lies in water depths ranging from 360 to 380 m. The semi-submersible design is a ring pontoon type of structure with four columns (Figure~\ref{fig:GjoaSemi}). The mooring system features 16 mooring lines arranged in clusters of 4.

The empirical meta-model aims at representing the relationship between the environment (wave, wind and current) and the tension in a mooring line. It is fitted on the \emph{in-situ} measurements of the environment and of the response of the structure. The waves and current have been measured by a MIROS radar. The significant wave height ($H_s$), wind speed ($W_s$) and current speed ($C_s$) and the corresponding directions are used. The tensions in the 16 mooring lines are sampled at 1 Hz. 28 storms, between 2011 and 2016, 12 hours each, are selected with $H_s$ up to 13 m. Each 12 hours has been split into 20-min sea-states. The tension considered by the meta-model is the maximum tension in one of the mooring lines during a 20-min sea-state.

\subsection{Mooring line tension}
\begin{figure}[ht]
	\centering
		\includegraphics[trim={0.4cm 0 0.9cm 0},width=\columnwidth]{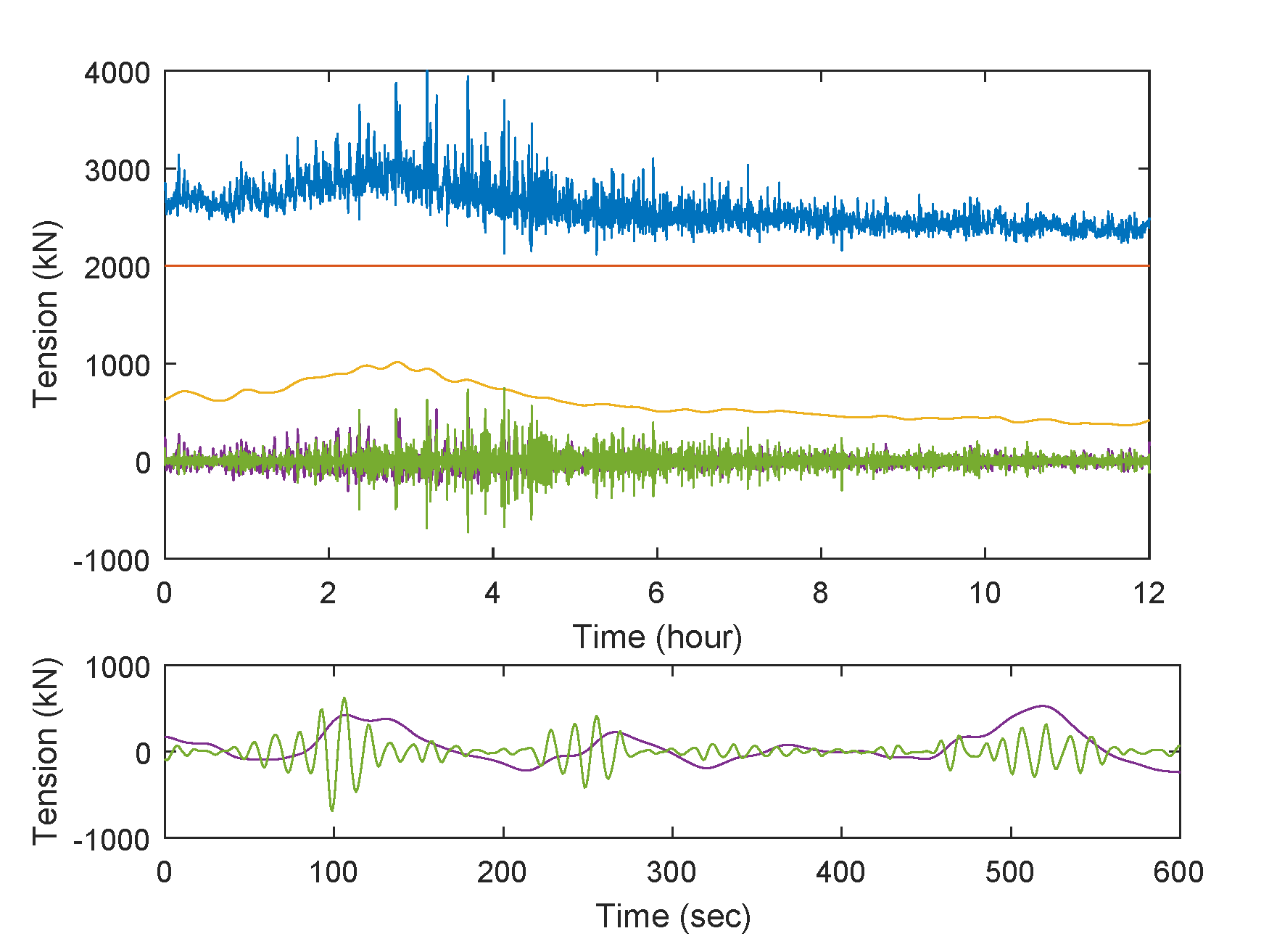}
	\caption{Decomposition of the tension in components - Total tension (blue), Pretension (red), Quasi-static (orange), Low-frequency (purple), High-frequency (green)}
	\label{fig:decomp_tension}
\end{figure}

The total tension in a mooring line comes from four different contributions: a static pretension, a quasi-static tension, a low-frequency dynamic tension and a high frequency dynamic tension. The decomposition is illustrated on Figure~\ref{fig:decomp_tension} (on the top 12 hours, on the bottom a zoom on 10 minutes of the LF and HF dynamic tensions).

\begin{itemize}
\item The pretension is the tension in the mooring which exists in the line when the structure is at rest, with no waves, wind and current. The pretension is supposed constant during all the 28 storms ($\sim$2000 kN).
\item The quasi-static tension (up to 1000 kN) is due to drift forces due to second order wave loads, and to wind and current (orange line in Figure~\ref{fig:decomp_tension}). It depends on the $H_s$, $W_s$ and $C_s$ and the corresponding directions and in a certain way on the mean wave frequency.
\item The low-frequency dynamic tension (up to 500 kN) is induced by the low frequency horizontal movements of the structure ($\sim$145 sec), mainly surge and sway, themselves generated by second order low frequency wave loads (purple lines in Figure~\ref{fig:decomp_tension}). It depends on $H_s$, not really on the wave direction as sway and surge behaviour are very similar (the structure is more or less symmetrical), and depends on mean wave frequency and on frequency bandwidth.
\item The high-frequency dynamic tension (up to 800 kN) is due to the pitch, roll and heave movements of the platform, linear platform responses to the wave kinematics (green lines in Figure~\ref{fig:decomp_tension}). It depends on $H_s$, not really on the wave direction as the structure is more or less symmetrical, and depends on mean wave frequency.
\end{itemize}

As the frequency information of waves is not available to the project, the meta-model is constructed considering only $H_s$, $W_s$ and $C_s$ and the corresponding directions.

\subsection{Estimation of the meta-model parameters}

On each 20-min sea-state, we first calculate $H_s$, wind and current mean speeds, and the corresponding mean directions. Then a Fourier band-pass filtering separates the tension components. Then the mean quasi-static tension, standard deviation and maximum of LF and HF tensions are calculated. The parameters of the following models are estimated by a least square method.

\subsubsection{Quasi-static tension}

In one part the drift forces are proportional to the square of the amplitude of the environmental loadings and in another part the structure is approximately symmetrical in the direction of the mooring line considered which is 225\degree, then the model for the quasi-static tension is taken as:

\begin{equation}
\begin{aligned}
T_{qs} &= \alpha_H H_s^2 (\cos(D_m)+\sin(D_m)) \\
       &\quad +\alpha_W W_s^2 (\cos(\Wdirm)+\sin(\Wdirm)) \\
       &\quad +\alpha_C C_s^2 (\cos(\Cdirm)+\sin(\Cdirm))
\label{Tqs}
\end{aligned}
\end{equation}

\subsubsection{LF dynamic tension}
As a second order effect, the standard deviation of the LF tension is related to the square of the $H_s$, and as told previously independent of the wave direction. As the behaviour of the mooring line is modified by the quasi-static tension, this tension has also been introduced in the meta-model.
The model for $\sigma_{LF}$, the standard deviation of the LF tension is then:

\begin{equation}
\sigma_{LF} = a_{LF} H_s^2+b_{LF} T_{qs} |T_{qs}|
\label{sLF}
\end{equation}

\subsubsection{HF dynamic tension}
The standard deviation of the HF tension is mainly proportional to $H_s$, and as told previously independent of the wave direction. As the behaviour of the mooring line is modified by the quasi-static and LF dynamic tensions, these tensions have also been introduced in the meta-model. The model for $\sigma_{HF}$, the standard deviation of the HF tension is then:

\begin{equation}
\sigma_{HF} = a_{HF} H_s+b_{HF} H_s^3+c_{HF} T_{qs} |T_{qs} |+d_{HF} \sigma_{LF}^2
\label{sHF}
\end{equation}

\subsubsection{Relation between standard deviation and 20-min maximum}

\begin{figure}[ht]
	\centering
		\includegraphics[width=\columnwidth]{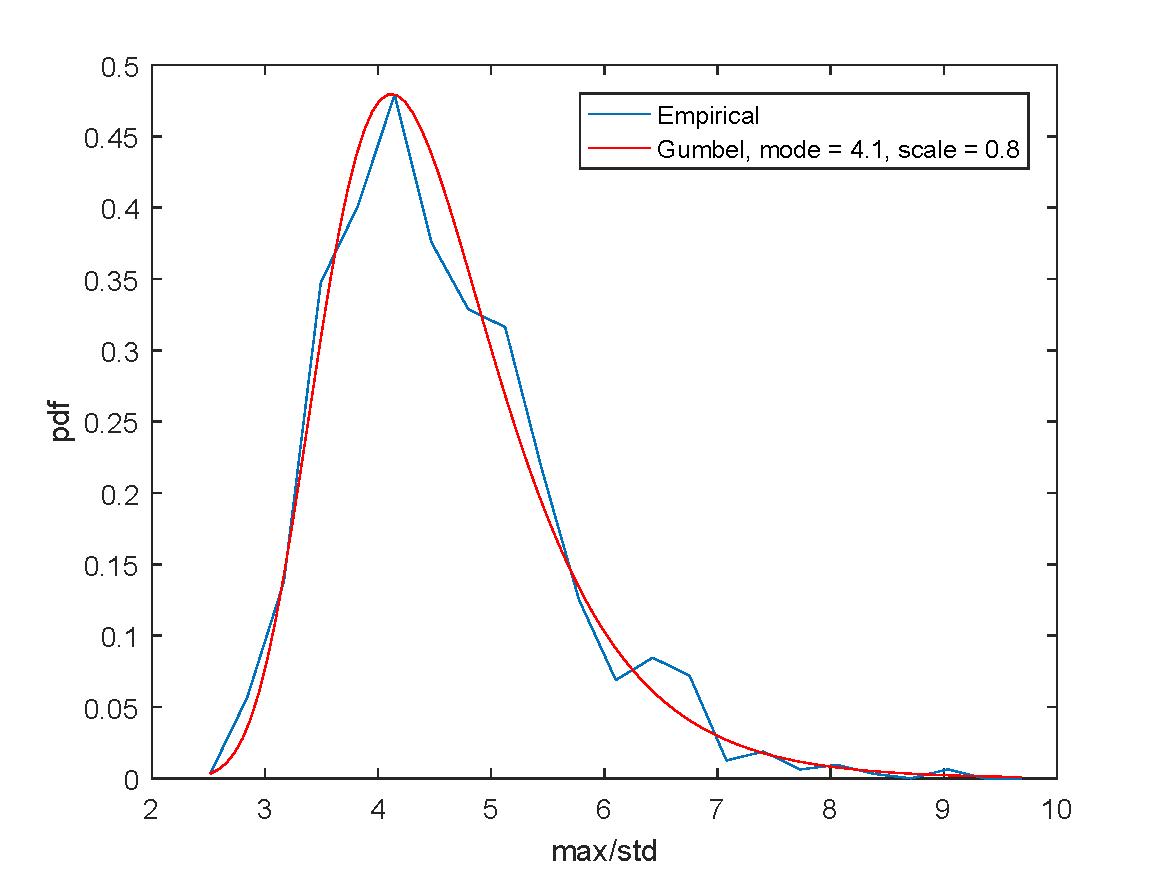}
	\caption{Fitting of a Gumbel distribution on the density of the LF tension maxima}
	\label{fig:fit_gumb}
\end{figure}

The statistical distribution of the maximum value of a random process on a duration $D$ tends (when $D$ is large) under some restrictive hypotheses to a $\GEV$ (Generalized Extreme Value) distribution (see \cite{Coles2001}). A particular case of this distribution is the Gumbel distribution. We have fitted this distribution on the empirical distributions of the normalized 20-min maxima of LF and HF dynamic tensions. For example, for the HF tension the distribution has the form:

\begin{equation}
P(T_{HF}^{max}/\sigma_{HF} \leq r)=\exp\left(-\exp\left(-\frac{r-\mu_{HF}}{\beta_{HF}}\right)\right)
\label{PGumbel}
\end{equation}

where $T_{HF}^{max}$ is the 20-min maximum HF tension, $\sigma_{HF}$ the standard deviation of the HF dynamic tension, $\mu_{HF}$ and $\beta_{HF}$ respectively the mode and the scale parameter of the Gumbel distribution. {The fittings were very good as shown in Figure~\ref{fig:fit_gumb} for the LF tension. The quality is the same for HF tension.}
The value of the maximum that we will consider could be the most probable value or a higher quantile. For a good fit with the measurements we have chosen $r_{HF}^{75\%}$ the quantile $75\%$, defined by Eq.~\eqref{PGumbel} and

\begin{equation}
P\left(T_{HF}^{max}/\sigma_{HF} \leq r_{HF}^{75\%} \right)=0.75
\label{PTHF}
\end{equation}

\subsection{Meta-model}
The meta-model for the maximum tension is constructed from Eqs~(\ref{Tqs}-\ref{PTHF}), with $T_{pre}$ the pretension.

\begin{equation}
T_{max} = T_{pre}+T_{qs}+r_{LF}^{75\%} \sigma_{LF}+r_{HF}^{75\%} \sigma_{HF}
\label{Tmax}
\end{equation}

\begin{figure}[ht]
	\centering
		\includegraphics[width=\columnwidth]{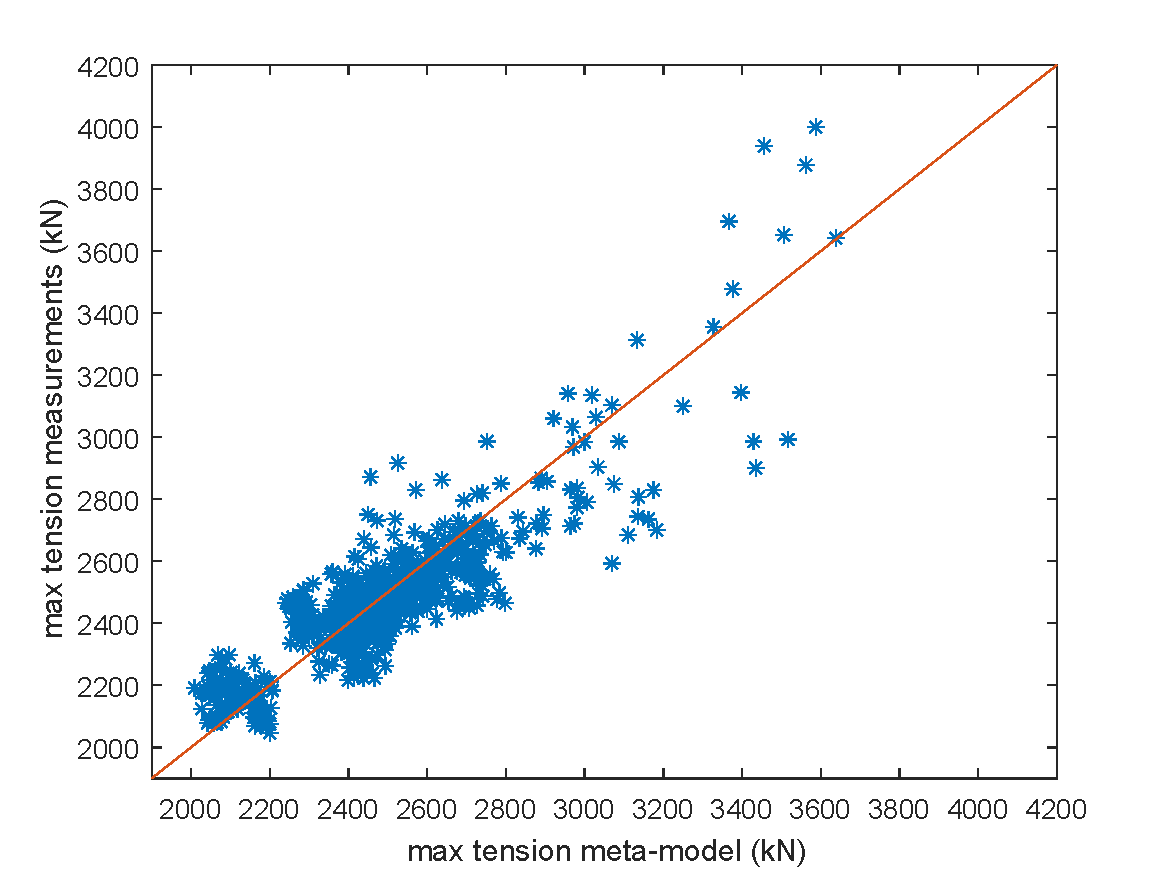}
	\caption{Comparison maximum tension meta-model vs measurements}
	\label{fig:meta_vs_meas}
\end{figure}

The appropriateness of the model to the measurements is shown in Figure~\ref{fig:meta_vs_meas}, which compares the total maximum tension measured in the mooring line to the value given by the meta-model Eq.~\eqref{Tmax}.

\section{Environmental dataset}

\subsection{Available data}

As explained before, the meta-model is constructed from \emph{in-situ} measurements, although for practical applications of structural design such data are not available. It is of common practice for structural design to rely on hindcast databases (see e.g. \cite{Jonathan2013}). Hindcast databases are based on numerical models and hence are available on a regular time step, over a fine spatial grid, in many regions of the world. The Ifremer IOWAGA wave database (see \cite{Rascle2013}), which is a recently developed model that provides accurate estimation of waves based on WAVEWATCH III\textregistered\, has been used to provide long time series of $H_s$. The wind from the NCEP CFSR \cite{Saha2010} reanalysis dataset is used as forcing of the numerical wave model, and is also used here for wind time series. The global HYCOM reanalysis \cite{Cummings2005} is used for estimation of currents. No circulation model forced by NCEP CFSR in North sea was available for this work. HYCOM was chosen for the availability of 3-hourly surface current speed over a long period (from 1993).  The concomitant use of three databases used for metocean environment provide a consistent hourly dataset, see e.g. \cite{ailliot2011}. Thus, the simultaneous use of these hindcasts is relevant for combining their independent physical effects on offshore structures. For each dataset, we used the nearest neighbour to the semi-submersible location.

The resulting database spans from 1st January 1993 to 31st December 2015, with a hourly time-step. The sea-state is described by 8 parameters, detailed in Table \ref{tab_param}. { Since the platform is almost symmetric, hence not very sensitive to the directionality, only the} mean direction is considered for each variable (\textit{e.g.} the mean of $(H_s \cos(D_m),H_s \sin(D_m))$ for significant wave height).

\begin{figure*}[ht]
	\centering
		\includegraphics[width=\textwidth]{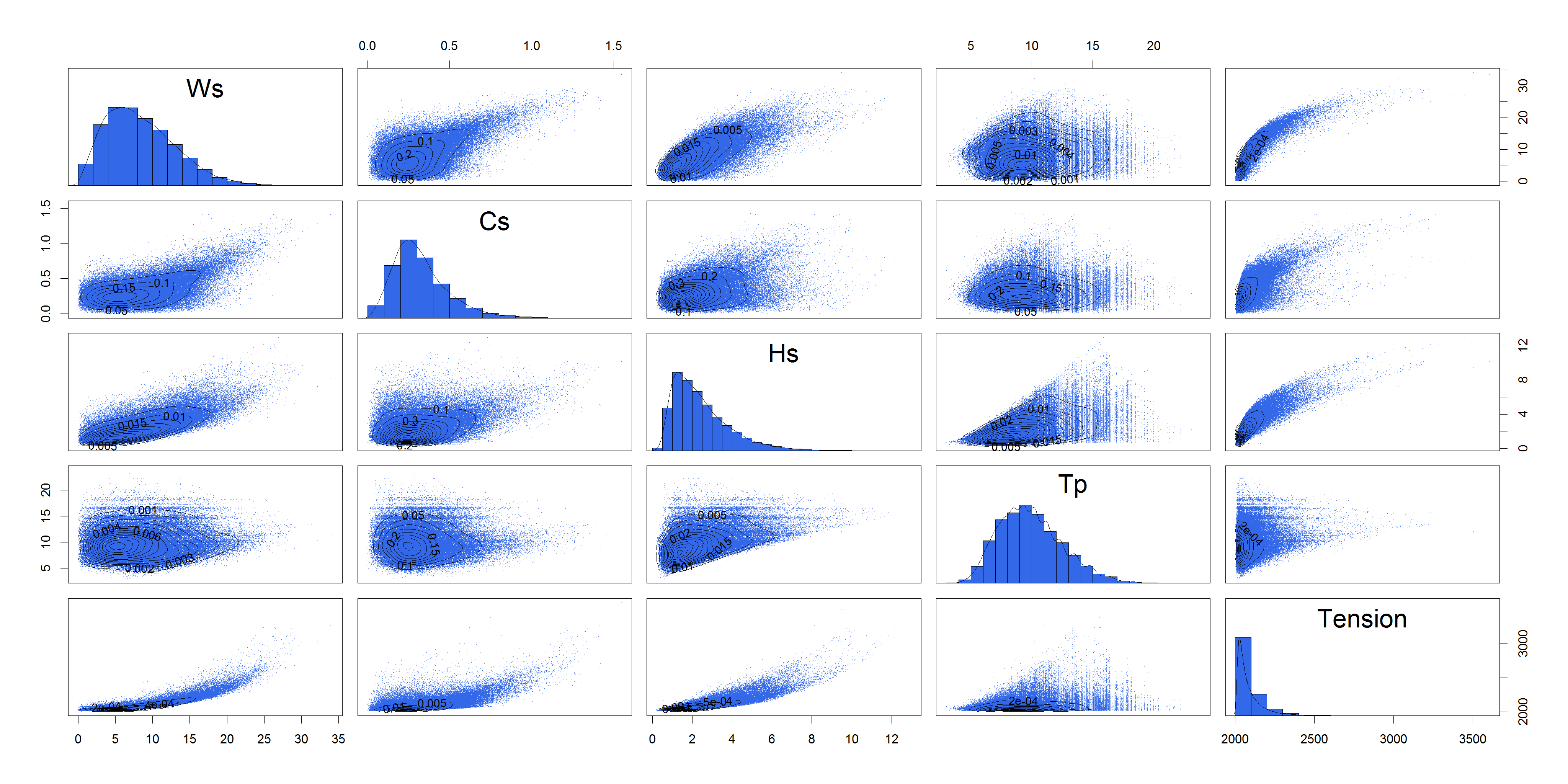}
	\caption{Descriptive statistics of sea-state parameters and model response}
	\label{fig:StatDesc}
\end{figure*}

The marginal and bivariate distributions of the four environmental variables can be seen in Figure~\ref{fig:StatDesc}, along with a kernel density estimation. Concerning the instantaneous distributions of the variables, one can clearly see the dependency between strong $H_s$ and strong $W_s$, between strong $W_s$ and strong $C_s$. On the other hand the extremal dependency between $H_s$ and $C_s$ is less obvious, with high wave values occurring whatever the value of the current. Concerning the tension in the mooring lines, one clearly sees the impact of wind and waves, and that the effect of the current is less important. This plot also points out that the marginal distribution of the tension has heavier tail that any of the other environmental parameters.


\subsection{Data analysis} \label{ssec:dataAnalysis}

\begin{figure*}[ht]
	\centering
		\includegraphics[width=\textwidth]{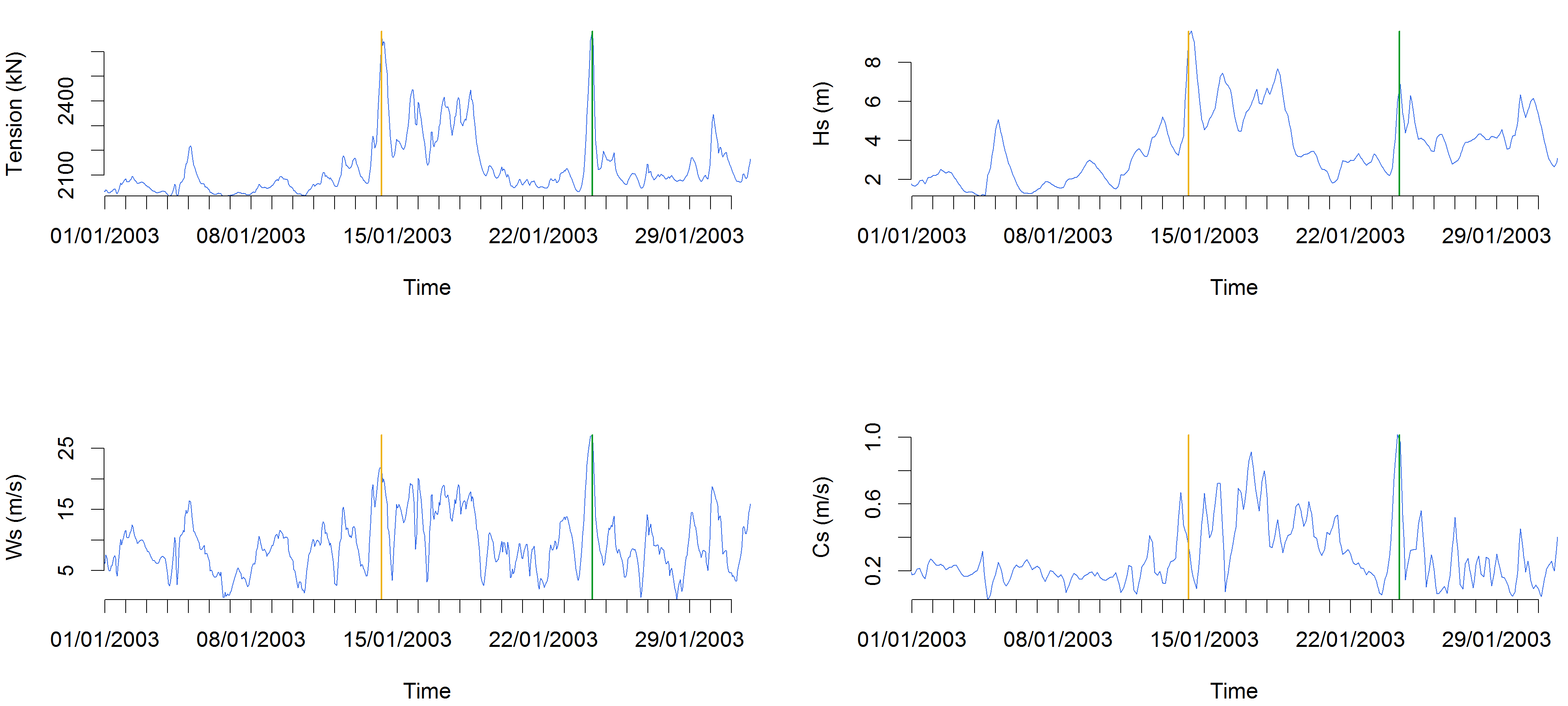}
	\caption{Example of synthetic response (upper left corner) and associated environmental conditions. See section \ref{ssec:dataAnalysis} for details.}
	\label{fig:Series}
\end{figure*}

Before going into the core of the methodology, a more precise insight into the data is provided in Figure~\ref{fig:Series}. This figure presents an extract of the full dataset during January 2003, where two important events of tension in the mooring lines can be observed. First in the upper-left corner, one can see the time series of the tension in a mooring line, and the corresponding environmental forcings in the other plots : $H_s$ in the upper left, $W_s$ in the bottom left and $C_s$ in the bottom right plot. Two vertical lines are superimposed on the plot, corresponding to extreme tensions. The left one, in yellow, corresponds to an extreme in the tension caused by conjunction of high wave height and strong wind, none of them being at its maximum as it can be seen from the plot : the $H_s$ is still increasing while the wind is decreasing. The right line, in green, corresponds to an extreme in the tension caused by a strong current, associated with strong wind, while the significant wave height is not so important (about 6 m).

The plot emphasizes the need for a joint modelling of the simultaneous distribution of the environmental parameters, because the joint occurrences of the different conditions leading to extreme response need to be quantified precisely. This will be the point of the following section.

\section{Methodology}

\subsection{Extreme values modelling}

\subsubsection{Univariate modelling}

As stated in the description of the data, since we deal with time-series with a time step small compared to the usual duration of storms, we need to retain only the maximal value within each storm before fitting an extreme model. This is done with the run-length method (see \cite{Ferro2003}), in which we define a cluster as the consecutive values exceeding a \emph{storm threshold}, $u_s$, and allowing the process to be under this threshold for some time $d_s$. If two clusters are separated from less that $d_s$ time steps, they are considered as only one cluster. Then, one takes the maximum value within each cluster. We thus obtain a sample of cluster maxima $X=(X_1,...X_{n_s})$, where $n_s$ is the number of clusters, or storms. { This method is called the run-length approach, see \cite{Coles2001}}. Following classical arguments, see e.g. \cite{Coles2001}, one can assume, that for a given sufficiently high threshold $u_\text{GPD}$, the distribution of $X_i$ conditionally on being above the threshold $u_\text{GPD}$ follows a \emph{Generalized Pareto Distribution} ($\GPD$) with parameters $\sigma>0$ and $\xi \in \mathds{R}$, best described by its cumulative distribution function (cdf) :
\[
\P(X<x | X > u_\text{GPD}) = 1-\left(1+ \xi \frac{x-u_s}{\sigma} \right)_+^{-1/\xi},
\] with $(z)_+ = \max(z,0)$.

We have chosen to estimate the parameters of the $\GPD$ { through} a maximum likelihood method. Results are shown in Table~\ref{tab:param.uvpot}. We have two parameters to fix here, $u_s$ and $u_\text{GPD}$. They are respectively fixed as the 97.5\% quantile and the 99\% quantile of the original dataset. Only the second one is reported since it impacts the estimated parameters. The duration $d_s$ separating two clusters has been fixed to 48 hours. We also checked (not shown here) that these values correspond to sensible choices by using the classical tools, such as residual life plot, the stability of estimates above the threshold, and the validity of the Poisson process hypothesis for threshold excesses, using the Dispersion Index plot {(\cite{Cunnane1979})}.

{ It can be seen from this table that the estimated shape parameter is positive only for the tension in the mooring lines, which is in accordance with the discussion of Figure~\ref{fig:StatDesc} and corresponds to a larger tail than the ones of the environmental variables.} 

\begin{figure}[ht]
	\centering
		\includegraphics[width=\columnwidth]{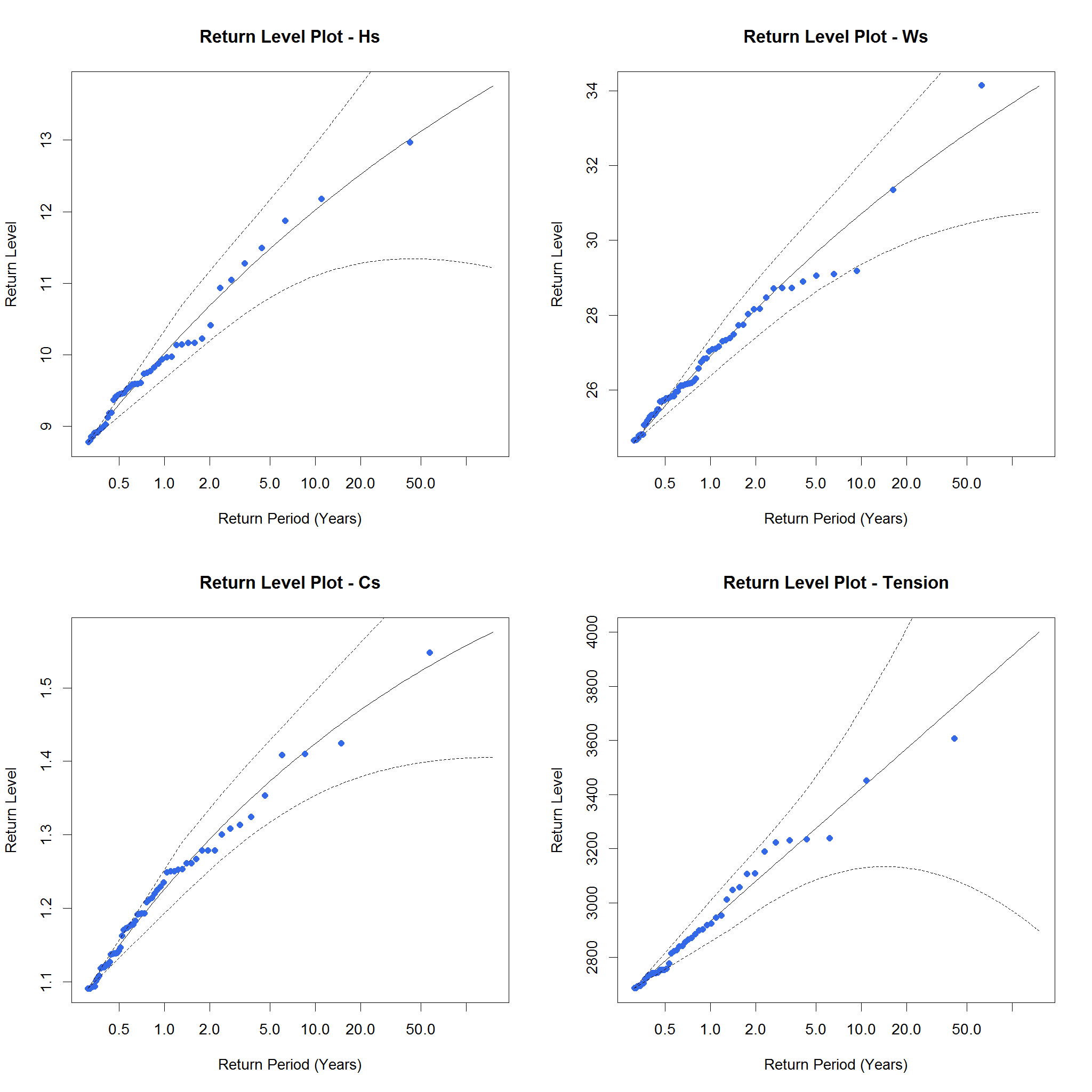}
	\caption{Return level plot of the fitted { POT} models}
	\label{fig:DiagFit}
\end{figure}

In Figure \ref{fig:DiagFit}, we show the return level plots for each variable. Each plot contains both the data (blue dots), the fitted models (solid black lines) and the corresponding confidence intervals computed using the delta method (dotted lines). It can be seen that the adjustment is very good for every variable and that the threshold used to fit the models is rather good.

The { 100-year return level} reported in Table~\ref{tab:param.uvpot} for the tension is the reference value, which would not be available for designing a new structure. Hence, our objective is to propose a method for deriving this value from the environmental data and from the knowledge of the meta-model.

\begin{table}[ht]
\centering
\begin{tabular}{rrrrrr}
  \toprule
 & Thres. & $n_s$ & $ \widehat{\sigma} \ \ $ & $\widehat{\xi} \ \ $ & Ret. level \\ 
  \midrule
$T_{\text{max}}$ & 2487.54 & 272 & 164.90 & 0.07 & 3825.35 \\ 
  $H_s$ & 7.13 & 210 & 1.59 & -0.16 & 13.35 \\ 
  $W_s$ & 21.53 & 359 & 2.70 & -0.13 & 33.69 \\ 
  $C_s$ & 0.89 & 345 & 0.17 & -0.18 & 1.55 \\ 
   \bottomrule
\end{tabular}
\caption{Parameters of the fitted GPD and associated return levels}
\label{tab:param.uvpot}  
\end{table}

\subsubsection{Joint extreme values modelling}

As stated in Section~\ref{MetaModel}, the structure is sensitive to three intensity variables, { meaning that the joint extremal structure of the environmental variables is needed in order to be able to compute extreme responses of the structure. Indeed,} it is quite straightforward to guess that taking the three 100-year return levels simultaneously in the meta-model (called \emph{perfect dependence} in {the following}) would lead to over-estimating the centennial return level for the tension, which is confirmed by the corresponding row in Table~\ref{tab:resRL}. Hence, the need for joint modelling of extreme is crucial, and some different models will be studied here.

In the 2D case, the standards and guidelines written by DNV-GL and IEC give some recommendations in the use of conditional laws. For instance, the log-normal model is recommended for $T_p$ given $H_s$ and for $H_s$ conditional on wind speed \cite{IEC2009}; the 2-parameter Weibull law is recommended for wind speed conditional on $H_s$ in North Sea \cite{DNV97}. To our knowledge, such recommendations do not exist for more than two variables. From a statistical point of view, these models are fitted to the bulk of the distribution, and hence may not be well suited for extreme values. Moreover, parametric models chosen to link the parameters of the distribution depending on the conditioning variable value are difficult to address, which may lead to misestimation of the true 100-year sea state.

In the sequel, we will focus on classical models that are able to deal with arbitrary dimensions. More precisely, let assume that our observations are the sample of cluster maxima, for which at least one component is extreme : $\bm{X} = (X_1,\ldots,X_d)$, a vector of $\R^d$ with GPD margins $F_i$, where $F_i$ is a $\GPD$ distribution with parameters $(\sigma_i, \xi_i)$. Then, the methods studied are the following :

\begin{itemize}
\item Independence : the components of $\bm{X}$ are assumed to be independent, hence its p.d.f is just the product of the $F_i$, which is a non-conservative approach ;
 
\item Perfect dependence : in this case, it is assumed that each return level occurs simultaneously, which is obviously conservative. In practice, it is usual to use for example the 100-year return level for waves and wind, coupled with the 20-year return level for current, without theoretical justification~;

\item Modified Nataf transform : the vector $\bm{X}$ is transformed to standard normally distributed marginals, and then one assumes that this vector is a Gaussian vector. The correlation matrix of the Gaussian vector is { chosen such that the  such that the correlation matrix of the data $\bm{X}$ is respected}. This approach is usual in structural safety, see e.g. \cite{Renard2007} for more details. In this study we modified the standard method to take better into account the tail of the joint distributions. The Gaussian correlation matrix is estimated such that the correlations between the components of the Gaussian vector above the threshold $u_\text{GPD}$ equals the correlations between the components of the data above the same quantile ratio~;

\item Logistic dependence function : $\bm{X}$ is transformed to a Fre\'chet scale vector $\bm{Z} =(Z_1,\ldots,Z_d)$ with $Z_i = _\frac{1}{\log F_i(X_i)}$. Its d-dimensional distribution function is taken to be $G(z_1,\ldots,z_n) = \exp[-V(z_1,\ldots,z_n)]$ where $V(z_1,z_2) = (z_1^{-1/\alpha}+z_2^{-1/\alpha})^{\alpha}$, $\alpha \in (0,1]$. This functional form is called \emph{Logistic dependence}, and can be extended to allowing asymmetry between the variables. The estimation has been carried out using a censored likelihood approach (see e.g. \cite{Raillard2014} for more details)~;

\item Conditional extremes : this approach, also referred to as Heffernan \& Tawn model, is a rather new and flexible semi-parametric approach to extremal dependence modelling. More precisely, a non-linear regression model fitted on each marginal of $\bm{Z}$ and assuming~: 
\[
\bm{Z}_{-i}|Z_i = \bm{a}_{-i|i} Z_i + Z_i^{\bm{b}_{-i|i}}\bm{\epsilon}_{-i|i} \text{, for } Z_i>\nu \text{ and } Z_{i} > \bm{Z}_{-i},
\]
where $\bm{Z}_{-i}$ if the vector of all variables, excluding $Z_i$ ;	$\bm{a}_{-i|i} \in [0,1]$ and $\bm{b}_{-i|i} \in (-\infty,1)$ are parameters of the fitted regression model ; $\nu$ is a { dependence} threshold above which the model is fitted and	$\bm{\epsilon}_{-i|i}$ are i.i.d with normal distribution $\N(\bm{\mu}_{-i|i},\bm{\sigma}_{-i|i})$, with the latter parameters $\bm{\mu}_{-i|i},\bm{\sigma}_{-i|i}$ estimated from using Maximum Likelihood. For more details, one can refer to \cite{Heffernan2004}.
\end{itemize}

{It could have been interesting to study other dependence structure, especially non-symmetric dependence functions. However, such a systematic study was beyond the scope of this project, which was to propose a clear methodology for extreme value modelling in a multivariate context. In addition, the study here covers all the possible dependence cases, from extremal dependence to extremal independence.}

\subsection{Environmental contours}

Here, we aim at comparing the <<response based>> and <<response independent>> approaches, based on the simplified meta-model presented in Section~\ref{MetaModel}. More precisely, we use the Huseby approach extended in 3D (as in \cite{Vanem2018}) to compute the environmental contours, and then the meta-model to calculate the response of the structure on this contour.

As explained before, the meta-model depends on three intensity parameters, (significant wave height, wind and current speeds) and the response was computed on the whole environmental data set, with {these} three parameters. Next, we computed the contours for low levels (up to the 98\% quantile) using the Huseby method, extended here to the 3-D case. {The method used is similar to the one developed in \cite{Vanem2018}, although the authors were not aware of the method proposed by Vanem when starting the project.}

\begin{figure}[ht]
	\centering
		\includegraphics[width=\columnwidth]{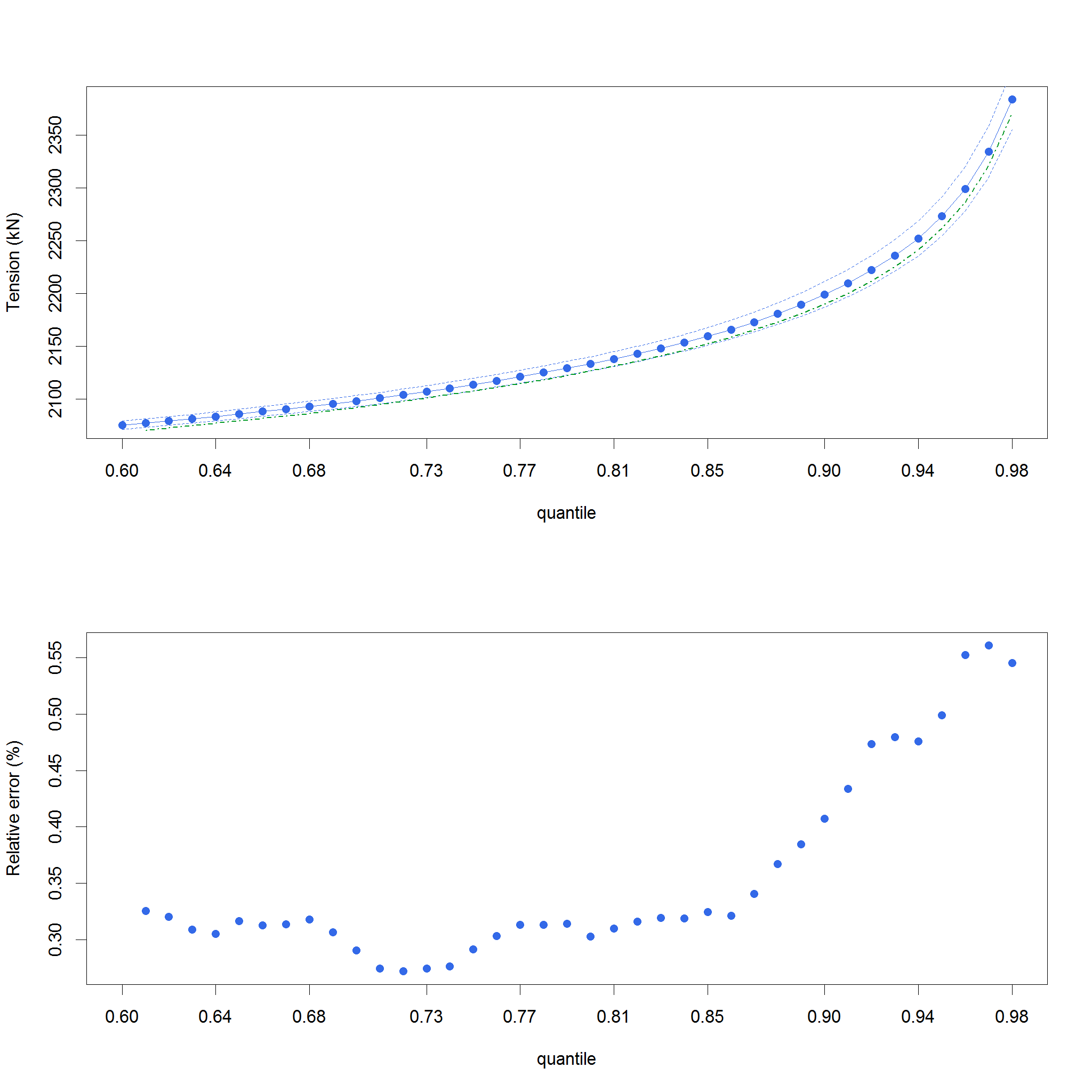}
	\caption{True response quantiles (blue dots) and 95\% confidence intervals (blue dashed lines), and quantiles estimated from the environmental contours (green dot-dashed line).}
	\label{fig:LowQuantiles}
\end{figure}

It can be seen from Figure~\ref{fig:LowQuantiles} that this method provides good estimates of the true response quantiles : the green curve, corresponding to the response estimated from the environmental contours, lies inside the 95\% confidence interval of the response quantiles estimated from the whole time series of responses. In the bottom plot of Figure~\ref{fig:LowQuantiles}, the relative error in estimating the response from the environmental contour is presented. {It can be seen from this plot that the error in estimating high quantiles of the response from environmental contours increases as the quantile  increases, which is quite sensible because of the paucity of data in the tail This plot emphasizes the need for modelling the tail of the joint distribution of environmental variables in order to obtain accurate estimations of high quantiles of responses}.

\begin{figure}[ht]
	\centering
		\includegraphics[width=\columnwidth]{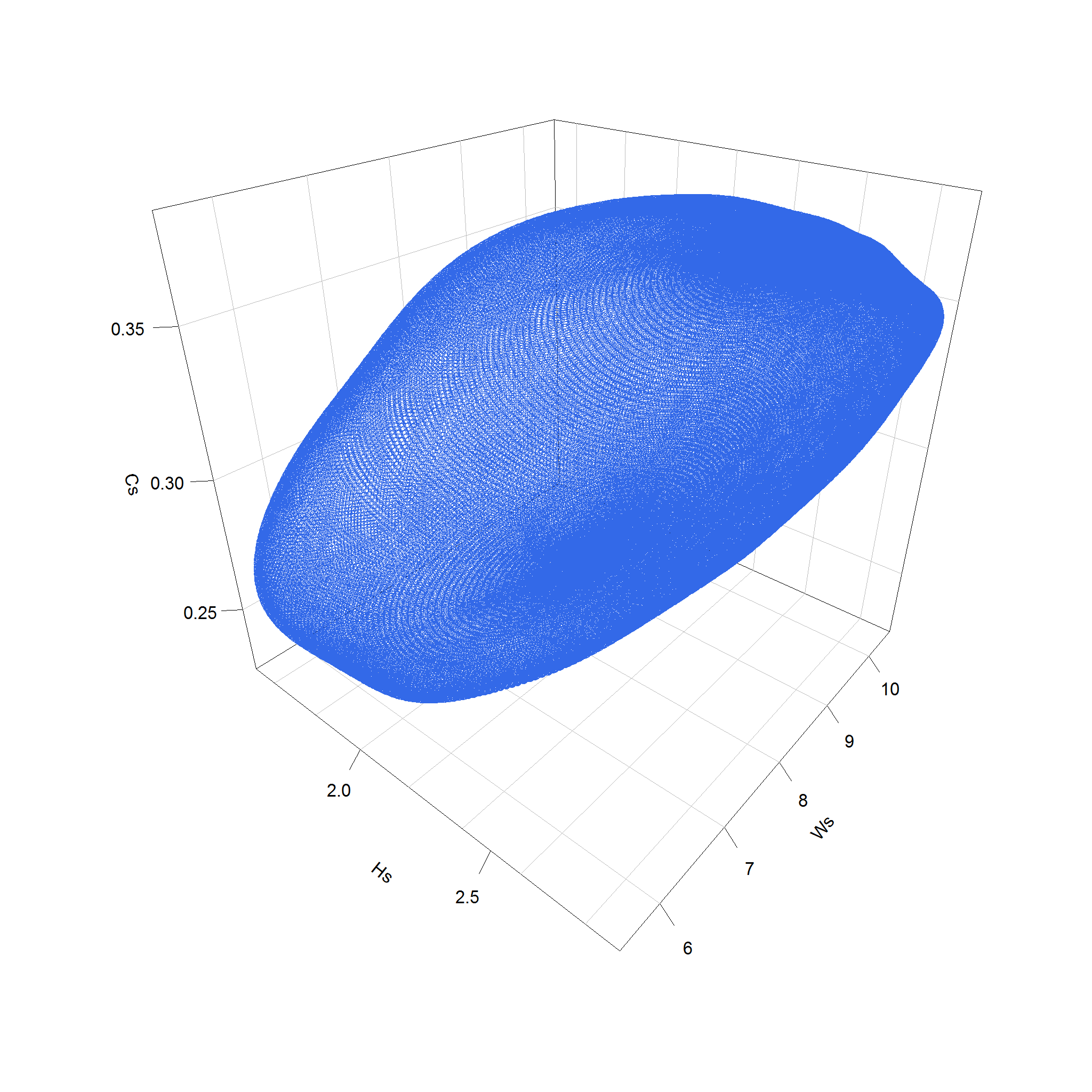}
	\caption{Environmental contour corresponding to a probability of exceeding of 0.3, {computed on the environmental database}.}
	\label{fig:LowContour}
\end{figure}

An example of a 3D contour is given in Figure \ref{fig:LowContour}. This plot shows each combination of $H_s$, $W_s$ and $C_s$ that is exceeded with a probability of 0.3. This empirical surface is here rather smooth, although such a surface could not be easily obtained for higher quantiles, due the the paucity of data in the distribution tails.

Once the environmental contour is defined, the tension is computed on the whole surface, and its maximum value is retained. This value is considered to be the 100-year return level, and the point at wich this value is maximal is called the Design point. Compared to computing the response over the whole space, the model needs only to be computed on a hyper-surface of the environmental parameters space.

\section{Comparison of results on extreme tensions in mooring lines}

In this section, we want to compare the aforementioned methods to compute extreme tensions in the mooring lines, namely the 100 years return level. The results are shown in Table~\ref{tab:resRL} for the estimation of the 100 year return level of the Tension, and in Table~\ref{tab:resDesignPoints} for associated design points. 

Our reference is the return level computed from the meta-model, which is first line in the table. As is can be seen, the perfect dependence, which is value obtained by assuming that the three 100-year values occur at the same time, greatly overestimates the tension: the value of 4273\ kN corresponds to the 300-year return level of the tension. In comparison, the two newly introduced methods provide values in accordance with the extreme tension computed for the meta-model. The 3-D Logistic model tends to slightly over-estimate the 100-year tension, while the conditional method slightly under-estimates it. This can be explained by the quite rigid structure imposed by the logistic model, which forces the three parameters to have a common dependence strength. The Modified Nataf transform also performs very well in this case, while it should be noticed that the Gaussian copula { is in the class of  asymptotic independence} copula and should thus be used with care in other applications.

\begin{table}[ht]
\centering
\begin{tabular}{lrr}
  \toprule
Method & Tension  & Relative Error \\
& (kN) & (\%) \\
  \midrule
Meta model & 3825 &  \\ 
  Independence & 3557 & -7\\ 
  Perfect dependence & 4273 & 12  \\ 
  Nataf transform & 3858 & 1 \\ 
  Logistic model & 4093 & 7 \\ 
  Conditional extremes & 3740 & -2\\ 
   \bottomrule
\end{tabular}
\caption{Comparison of the methods for the estimation of the 100-year return level.} 
\label{tab:resRL}
\end{table}

\begin{table}[ht]
\centering
\begin{tabular}{lrrrrr}
  \toprule
Method & $H_s$ (m) & $W_s$ (m/s) & $C_s$ (m/s)\\ 
  \midrule
  Independence  & 13.10 & 25.55 & 1.09 \\ 
  Perfect dependence & 13.35 & 33.69 & 1.55 \\ 
  Nataf transform & 13.16 & 31.37 & 1.20 \\ 
  Logistic model & 13.09 & 32.43 & 1.50 \\ 
  Conditional extremes & 11.92 & 31.97 & 1.40 \\  
   \bottomrule
\end{tabular}
\caption{Comparison of the design points.}
\label{tab:resDesignPoints}
\end{table}

\begin{figure*}[ht]
	\centering
		\includegraphics[width=\textwidth]{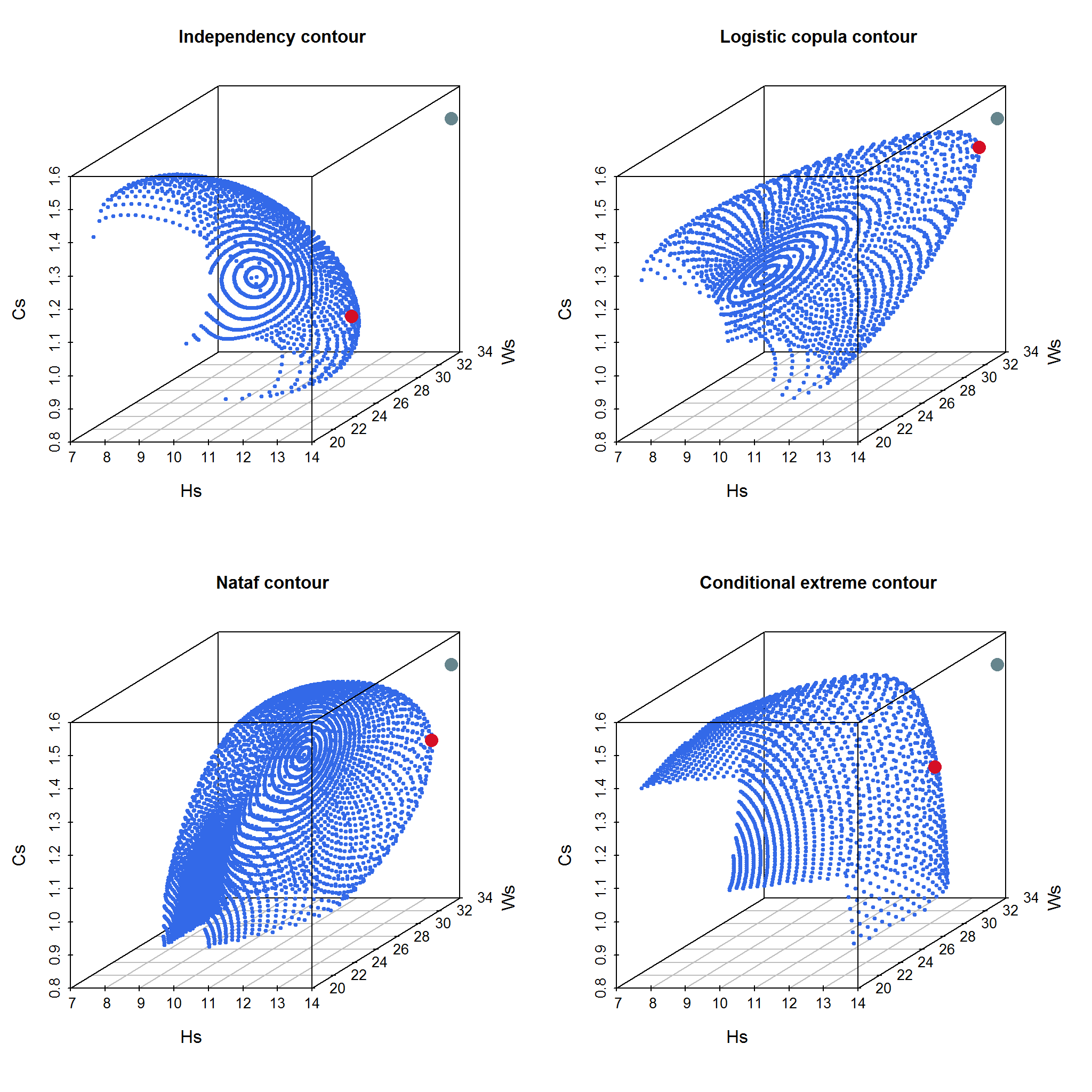}
	\caption{Environmental contours and design point (red). In grey, the perfectly dependent design point.}
	\label{fig:Contour}
\end{figure*}

If one compares the design values associated to the the return level computed from each method presented in Table~\ref{tab:resDesignPoints}, it can be seen that even if the current speed value is more or less the same, the design wind speed and design significant wave height are rather different depending on the method used. Usually, the contours are provided as curves in the plane to allow easier comparison of the methods used. Here, the contours are 3-D surfaces, and can be found in Figure~\ref{fig:Contour}. In the figure, one can see in each sub-figure the contour derived with the indicated method with blue dots. The red dots correspond to the design points, i.e. the point of the surface for which the response computed with the meta-model is the highest. In each plot is also represented the { perfectly} dependant case, with a grey dot.  

The differences in the dependence structure are stringent in this figure, since in the margins are the same for each case. Here, the conditional extreme model seems to be left-cut, which is caused by the modelling procedure: this model tackles the case where at least one component is extreme, while the others consider that all the components are above the threshold. This figure also emphasizes the dependence of the structure on the $H_s$ parameter, since the design points are all located along this direction.

It can be noticed from this plot that the Logistic copula leads to a dependence structure that is close to perfect dependence, which would be a straight line connecting the origin to the grey dot. This might be due to the symmetry imposed by the model, since all the variables are exchangeable in the model, while the { semi-parametric} model of Heffernan \& Tawn suggests that the dependence is stronger between $H_s$ and $W_s$ than the others. Extensions of the Logistic copula exist, but become rapidly intractable as the dimension grows. The use of parametric models taking into account asymmetry in the variables is left for future work, and one can conclude that the conditional extreme model is able to give very accurate prediction of the extreme level in the case considered here.

\section{Conclusion}

This paper presented some work achieved during the CITEPH project MulanR, whose objective was to explore new methodologies to ease the design phase of offshore structure. First, by constructing a meta-model of the structure, we were able to obtain a time-series of synthetic, yet realistic, response of the structure, and by then, an estimation of the centennial response. This value was then considered as a reference, in a <<response based>> approach. Then, we compared different classical and new methods to derive environmental contours, in a <<response independent>> approach, which leads in turns to an estimation of the centennial response of the structure, which can be compared to the reference obtained previously.

The key finding in this paper is the versatility of the Heffernan \& Tawn model, also referred to as conditional extreme model, which was found to be efficient in modelling 3-D extremes along with the use of Huseby's contouring approach. This approach was also found to be efficient for 2-D extremes, although it was not shown here, an interesting application can be found in \cite{Gouldby2017}. This result is shown for our case study and further investigations are needed for other extremal dependences, { although the adopted methodology is general enough to be adapted to other structures and locations (see \cite{Pineau2018} for examples e.g. including directionality as a discrete covariable).} However, this finding is important because it allows to extend the computation of design points above 2D, which is usually the case as far as offshore structure are concerned. However, attention should be paid on the difficulties associated with the threshold choice for this conditional extreme value model, but this question is left for future studies.

The good performance of the Modified Nataf model (Gaussian copula) can also be pointed out, given that this model is not an extreme value copula, and the question whether to consider it or not is left for future study.

\section*{Acknowledgements}

This research work has been carried out in the frame of the CITEPH project MulAnR, funded by both partnership and sponsors. The authors thanks ENGIE E\&P International, SAIPEM and Doris Engineering for their financial support and their fruitful discussions.

ENGIE E\&P Norge is acknowledged for the provision of storm time-series measured at Gj\o a platform.

The authors are also grateful to two anonymous referees for their valuable comments that leaded to great improvement of the paper.

\bibliographystyle{elsarticle-num-names}
\bibliography{biblio}

\end{document}